\documentclass[12pt]{article}

\usepackage{graphicx}
\usepackage{epstopdf}
\usepackage[centertags]{amsmath}
\usepackage{amssymb}
\usepackage{amsthm}
\usepackage{amsfonts}
\usepackage{ccaption}
\usepackage[usenames]{color}
\usepackage{comment}
\usepackage[square, comma, sort&compress]{natbib}

\usepackage[
      colorlinks=true,
      linkcolor=black,  
      urlcolor=blue,    
      filecolor=blue,  
      citecolor=black,   
      pdfstartview=FitV,
      pdftitle={},
        pdfauthor={Mukund Rangamani,Simon Ross,Dam T. Son,Ethan Thompson},
        pdfsubject={},
        pdfkeywords={},
        pdfpagemode=None,
        bookmarksopen=true    
      ]{hyperref}
       
\usepackage{rotating}

\makeatletter
\@addtoreset{equation}{section}

\makeatletter
\renewcommand\section{\@startsection {section}{1}{\z@}%
                                   {-3.5ex \@plus -1ex \@minus -.2ex}
                                   {2.3ex \@plus.2ex}%
                                   {\normalfont\large\bfseries}}
\renewcommand\subsection{\@startsection{subsection}{2}{\z@}%
                                     {-3.25ex\@plus -1ex \@minus -.2ex}%
                                     {1.5ex \@plus .2ex}%
                                     {\normalfont\bfseries}}

  \captionnamefont{\bfseries}
  \captiontitlefont{\small\sffamily}
  \captiondelim{: }
  \hangcaption

\def\baselinestretch{1.2}
\def\footnotesize{\@setsize\footnotesize{10pt}\xpt\@xpt
\abovedisplayskip 10\p@ plus2\p@ minus5\p@
\belowdisplayskip \abovedisplayskip
\abovedisplayshortskip  \z@ plus3\p@
\belowdisplayshortskip  6\p@ plus3\p@ minus3\p@
\def\@listi{\leftmargin\leftmargini
\topsep 6\p@ plus2\p@ minus2\p@\parsep 3\p@ plus2\p@ minus\p@
\itemsep \parsep}}
\long\def\@makefntext#1{\parindent 5pt\hsize\columnwidth\parskip0pt\relax
\def\strut{\vrule width0pt height0pt depth1.75pt\relax}%
$\m@th^{\@thefnmark}$#1}
\newcommand{\be}{\begin{equation}}
\newcommand{\ee}{\end{equation}}
\newcommand{\beq}{\begin{eqnarray}}
\newcommand{\eeq}{\end{eqnarray}}

\def\sec#1{\S \ref{#1}}

\def\req#1{(\ref{#1})}
\def\App#1{Appendix \ref{#1}}

\def\d{\delta}
\def\eps{\epsilon}

\def\r{\rho}

\def\thus{\Longrightarrow}

\def\CB{{\cal B}}

\def\CD{{ \cal D }}

\def\CH{{\cal H}}

\def\CM{{\cal M}}
\def\CN{{\cal N}}

\def\CQ{{\cal Q}}
\def\CR{{\cal R}}
\def\CS{{\cal S}}

\def\CX{{\cal X}}

\def\R{{\bf R}}
\def\Sp{{\bf S}}

\def\A5S5{{\rm AdS}_5 \times \S^5}

\def\AdS#1{{\rm AdS}_{#1}}
\def\Schr#1{{\rm Schr}_{#1}}
\def\q{\gamma^2}
\def\q{\gamma^2}
\def\vel{v}
\def\u{u}
\def\vsq{v^2}

\renewcommand\d{\partial}

\newcommand\etar{\eta_{\rm rel}}

\def\({\left (}
\def\){\right )}
\def\[{\left[}
\def\]{\right]}

\def\beq{\begin{equation}}
\def\eeq{\end{equation}}
\newcommand{\bea}{\begin{eqnarray}}
\newcommand{\eea}{\end{eqnarray}}
\def\er{\epsilon_{\rm rel}}
\def\pr{P_{\rm rel}}

\def\ef{j_\varepsilon}
\def\r{\rho}
\def\dt{\partial_t}
\def\di{\partial_i}
\def\dj{\partial_j}

\def\pnr{P}

\def\enr{\varepsilon}

\title{{\bf \Large Conformal non-relativistic hydrodynamics from gravity}}

\author{
Mukund Rangamani$^a$\footnote{mukund.rangamani@durham.ac.uk}, \ 
Simon F. Ross$^a$\footnote{s.f.ross@durham.ac.uk} , \ 
D. T. Son$^b$\footnote{son@phys.washington.edu}, \
Ethan G. Thompson$^c$\footnote{egthomps@u.washington.edu}\\ \\
\small \sl $^a$  Centre for Particle Theory \& Department of
Mathematical Sciences,
\\[-1.5mm]
\small \sl Science Laboratories, South Road, Durham DH1 3LE, United Kingdom. \\
\small\sl $^b$ Institute for Nuclear Theory, University of Washington, 
Seattle, WA, 98195-1550, USA \\
\small\sl $^c$Department of Physics, University of Washington, 
Seattle, WA, 98195-1560, USA
}

\begin{document}
\setlength{\baselineskip}{16pt}
\begin{titlepage}
\maketitle
\begin{picture}(0,0)(0,0)
\put(350, 300){DCPT-08/61} 
\put(350,285){INT PUB 08-48}
\end{picture}
\vspace{-36pt}

\begin{abstract}
We show that the recently constructed holographic duals of conformal
non-relativistic theories behave hydrodynamically at long distances, and 
construct the gravitational dual of fluid flows 
in a long-wavelength approximation.  
We compute the thermal conductivity of the holographic 
conformal non-relativistic fluid.  The 
corresponding Prandtl number is equal to one.
\end{abstract}
\thispagestyle{empty}
\setcounter{page}{0}
\end{titlepage}

\renewcommand{\baselinestretch}{1.4}  
\renewcommand{\thefootnote}{\arabic{footnote}}


\tableofcontents
\section{Introduction}
\label{intro}

The AdS/CFT correspondence \citep{Maldacena:1997re, Gubser:1998bc,
Witten:1998qj} provides an important theoretical framework for
studying a class of strongly coupled quantum field theories, and at
the same time provides a non-perturbative approach to quantum
gravity. Traditionally, the correspondence has been used to study the
dynamics of non-abelian gauge theories, since the prototype examples
of the correspondence relate string theory/gravity in AdS spacetimes
to supersymmetric Yang-Mills theories. However, in recent times the
correspondence has been extended to study various condensed matter
systems, especially those in the vicinity of a quantum critical
point (see \citep{Sachdev:2008ba} and references therein). 
An example is the possible application to non-relativistic
fermions at unitarity \citep{Son:2008ye,Balasubramanian:2008dm}, which
is the focus of the current paper.

Fermionic systems at unitarity, which are realized in experiments with
cold atoms, are described by a non-relativistic CFT. This field theory
enjoys a symmetry algebra called the Schr\"odinger algebra
\citep{Hagen:1972pd,Mehen:1999nd,Nishida:2007pj}.  The authors of
\citep{Son:2008ye,Balasubramanian:2008dm} proposed a class of
spacetimes which could describe field theories with non-relativistic
conformal invariance holographically, whose metric is 
\begin{equation}
ds^2 = r^2\, \left(-2 \, dx^+\, dx^- - \beta^2 \,r^{2}\, (dx^+)^2 
+ d{\bf x}_d^2 \right) + \frac{dr^2}{r^2}.
\label{vacmet}
\end{equation}	
These have the Schr\"odinger algebra as an isometry algebra, and we will
henceforth refer to them as Schr spacetimes. A Galilean CFT in
$d$-spatial dimensions is conjectured to be dual to $\Schr{d+3}$.

Subsequently, the $\Schr{5}$ spacetime was realized in string theory
\citep{Herzog:2008wg,Maldacena:2008wh,Adams:2008wt}.\footnote{For other recent studies of non-relativistic systems in the context of gauge-gravity duality see \citep{Goldberger:2008vg,Barbon:2008bg, Wen:2008hi, Nakayama:2008qm,Chen:2008ad,Minic:2008xa,Imeroni:2008cr,Galajinsky:2008ig,Kachru:2008yh,Pal:2008rf,SekharPal:2008uy,Pal:2008id,Kovtun:2008qy,Duval:2008jg,Yamada:2008if,Lin:2008pi,Hartnoll:2008rs,Schvellinger:2008bf,Mazzucato:2008tr,Leiva:2003kd}.} It was obtained by starting from a known type IIB solution, the $\AdS{5} \times \CX_5$
spacetime, where $\CX_5$ is a Sasaki-Einstein manifold, and applying a solution-generating transformation, either the Null Melvin Twist \citep{Alishahiha:2003ru,Gimon:2003xk} or the TsT transformation \citep{Lunin:2005jy}, to generate the spacetime $\Schr{5} \times \CX_5$. The TsT transformation involves a T-duality, followed by a shift, and then a second T-duality. This can be used in
any spacetime with a $U(1) \times U(1)$ isometry and, in general, can
be thought of as a Melvin Twist, since the transformation adds a
background NS-NS B-field, effectively Melvinizing the spacetime. The
Null Melvin Twist can be shown to be equivalent to the TsT
transformation in the special case when one of the $U(1)$ isometries
is null.

Since the string-theoretic embedding of the $\Schr{5}$ spacetimes can
be achieved by using dualities, we can see that the dual field theory
is a deformed version of $\CN =4$ SYM theory (for the special case of $\CX_5 = \Sp^5$). The transformation in \citep{Herzog:2008wg,Maldacena:2008wh,Adams:2008wt} breaks the $SU(4)$ symmetry of $\CN =4$ down to an $SU(3) \times U(1)$ subgroup, and the
field theory is then the theory twisted by the R-charge associated
with the $U(1)$ isometry.  One can also view the field theory as the
one obtained by decoupling the open string field theory living on
D3-branes in a Null Melvin spacetime. More generally any $\CN=1$ superconformal field theory in $3+1$ spacetime dimensions (corresponding to a dual Sasaki-Einstein compactification) can be deformed to a two dimensional (spatial)  non-relativistic CFT using the $U(1)_R$ symmetry.

Using the same solution-generating transformation, the authors of
\citep{Herzog:2008wg,Maldacena:2008wh,Adams:2008wt} constructed a
two-parameter family of black hole spacetimes with $\Schr{5}$
asymptotics. These black holes can be shown to solve the
five-dimensional effective equations of motion obtained by a
Kaluza-Klein reduction of the 10-dimensional type IIB theory on the
$\CX_5$. The simplest five dimensional effective action is composed of
gravity coupled to a massive vector field and a single scalar field
\citep{Herzog:2008wg}, which is a truncation of a more general
five-dimensional Lagrangian involving three scalar fields. The latter,
remarkably, is a consistent truncation of type IIB supergravity on
$\CX_5$ \citep{Maldacena:2008wh}.

This black hole solution was used to study the equilibrium
thermodynamic properties of the field theory and was shown to be dual
to the grand canonical ensemble for the dual field theory. The
thermodynamics is, not surprisingly, consistent with non-relativistic
scale invariance in two spatial dimensions.  In particular, it was
found that $\varepsilon=P$, where $\varepsilon$ is the energy density
and $P$ is the pressure, as required in non-relativistic CFTs.  Furthermore,
non-equilibrium transport properties of the non-relativistic plasma
were also explored in \citep{Herzog:2008wg,Adams:2008wt}. In
particular, the shear viscosity $\eta$ of the non-relativistic fluid
was calculated and found to take the universal value $\eta/s = 1/4\pi$
typical of strongly interacting field theories with gravity duals.

The present paper is concerned with constructing the gravitational
dual of arbitrary fluid flows in the non-relativistic CFT. 
The basic premise is to build on the recent ideas in the {\it
fluid-gravity correspondence} \citep{Bhattacharyya:2008jc} to construct
inhomogeneous black hole solutions with $\Schr{5}$ asymptotics.
The hydrodynamic description of a fluid (either relativistic or
non-relativistic) is an effective field theory which captures the
universal long wavelength physics when the system achieves local
thermal equilibrium.  Building upon previous discussions of the
hydrodynamic description of four-dimensional superconformal field
theories in the AdS/CFT correspondence (see \citep{Son:2007vk} and
references therein), in
\citep{Bhattacharyya:2008jc} it was argued that starting from the most
general black hole solution in AdS (a boosted Schwarzschild-AdS black
hole) one can promote the temperature and velocity fields to local
functions of the boundary coordinates. It is then possible to solve
for the bulk metric order by order in a boundary derivative expansion
and recover the relativistic Navier-Stokes equations, which encode the
conservation of the boundary energy-momentum tensor, entirely as a
consequence of Einstein's equations. Importantly, it is possible to
show that the resulting bulk solutions are genuine black holes, i.e.
that they have a regular event horizon
\citep{Bhattacharyya:2008xc}. This correspondence has been extended in
many directions, including to forced fluids
\citep{Bhattacharyya:2008ji}, to conformal field theories in various
dimensions \citep{VanRaamsdonk:2008fp,
Haack:2008cp,Bhattacharyya:2008mz}, to charged fluids
\citep{Erdmenger:2008rm,Banerjee:2008th,Hur:2008tq}, to Bjorken flow \cite{Heller:2008mb, Kinoshita:2008dq}, and incompressible non-relativistic fluids \cite{Bhattacharyya:2008kq}.

We are interested in constructing inhomogeneous black hole solutions
with Schr\"odinger asymptotics and extending the fluid-gravity
correspondence to relate them to fluid flows in the non-relativistic
conformal theory. As a first step, we note that a four parameter family
of black hole solutions with $\Schr{5}$ asymptotics can be obtained by
boosting the solutions considered in
\citep{Herzog:2008wg,Maldacena:2008wh,Adams:2008wt}. We can then
promote these parameters to functions of the field theory coordinates
and find the solution order by order in a boundary derivative expansion.
However, we will show that we can obtain inhomogeneous black hole
solutions with the desired asymptotics in an easier way, by simply
applying a TsT transformation to the solutions constructed in
\citep{Bhattacharyya:2008jc}! We will argue that this in fact captures
all the hydrodynamic properties of the system in the planar
limit. 

To relate these inhomogeneous black hole solutions to fluid flows, we
need to calculate the boundary stress tensor from the bulk
solutions. Because of the slow fall-off of the metric perturbation in
the asymptotically Schr\"odinger black holes, the usual technique of
obtaining the stress tensor by functionally differentiating the action
with respect to the boundary metric cannot be straightforwardly
applied to these cases (see \App{appHam} for a discussion of this issue). However, \citep{Maldacena:2008wh} proposed that
the stress tensor of the asymptotically AdS space before the TsT
transformation can be re-interpreted as the stress tensor complex of
the non-relativistic theory. We will adopt this approach. We therefore
describe the general reduction of a relativistic stress tensor on the
light cone to obtain a non-relativistic stress tensor complex. The
structure of the relativistic conformal stress tensor implies that for
any non-relativistic conformal theory obtained in this way, the
thermal conductivity $\kappa$ of the non-relativistic fluid is
\begin{equation}
\kappa = 2\,\eta\, \frac{\enr+P}{\rho\,  T}
\label{kappaval1}
\end{equation}
where $\rho$ is the mass density, or, even more succinctly, 
\begin{equation}
  \mathrm{Pr} = 1
\end{equation}
where $\mathrm{Pr}$ is the Prandtl number.

The outline of this paper is as follows: we will begin in \sec{lcredn}
by discussing how the relativistic fluid equations are reduced on the
light-cone to the non-relativistic Navier-Stokes equations. This will
allow us to explore the general properties of the non-relativistic
stress-tensor complex. We will review some aspects of the $\Schr5$
solutions in \sec{gravdisc} and then describe how to construct
inhomogeneous black hole solutions dual to arbitrary fluid flow and
give the dual stress tensor to first order in derivatives in
\sec{nrfluids}. We end with a discussion in \sec{discuss}.

\section{Light-cone reduction of relativistic fluids}
\label{lcredn}

Consider a relativistic fluid in Minkowski space in $d+2$ spacetime
dimensions; we will use light-cone coordinates $\{x^+,x^-,{\bf x}\}$
and take the metric to be
\begin{equation}
ds^2_{\rm flat} = \eta_{\mu \nu} \, dx^\mu \, dx^\nu 
  = -2 \, dx^+ \, dx^- + d{\bf x}^2 .
\label{flatlc}
\end{equation}	
Suppose we view this fluid in the light-cone frame and evolve it in
light-cone time $x^+$. Then, for fixed light-cone momentum $P_-$, we
obtain a system in $d+1$ dimensions with non-relativistic
invariance. This is of course familiar from the discrete light-cone
quantization (DLCQ) of quantum field theories. In fact, one of the
models for studying non-relativistic conformal field theories
holographically, suggested in refs.\
\citep{Goldberger:2008vg,Barbon:2008bg}, was that one could consider
pure AdS, with the relativistic conformal symmetry broken to Galilean
symmetry simply by compactification of the $x^-	$ coordinate, which
singles out a preferred light-cone direction. Note that in this case
we are not only compactifying the light-cone direction in the boundary
where gravity is non-dynamical (and the metric flat, \req{flatlc}), we
are also required to compactify the coordinate in the bulk AdS
spacetime. This involves introducing closed null curves in the
geometry and the validity of supergravity becomes questionable
\citep{Maldacena:2008wh}. We will return to a different gravitational
background, viz. \req{vacmet},  where $x^-$ does not need to be
compactified to achieve Galilean symmetry. Note however that it is
still useful to take $x^-$ compact, so that the momentum $P_-$ is
integer quantized, since $P_-$ is interpreted as particle number in the
dual theory. 

Relativistic hydrodynamics in $d+2$ dimensions is formulated in terms
of pressure (or, equivalently, the temperature) and the four velocity
$\u^\mu$, subject to the condition that $\eta_{\mu \nu} \, \u^\mu \,
\u^\nu = -1$.  This gives $d+2$ degrees of freedom. At the same time,
non-relativistic hydrodynamics in $d$ spatial, one temporal
dimensions can be formulated in terms of the mass density $\rho$, the
pressure $P$, and the spatial velocities $\vel^i$, also giving $d+2$
degrees of freedom.

We would like to find a mapping between the degrees of freedom of the
$(d+2)$-dimensional theory to the degrees of freedom of the $d+1$
dimensional theory such that the relativistic hydrodynamic equations
imply the non-relativistic hydrodynamic equations.  We would also like
to find how the thermodynamic quantities of the two formulations are
related. Finally, we plan to use the map to find out the thermal
conductivity of the non-relativistic theory. We will first begin with
ideal hydrodynamics and then discuss dissipative terms.

\subsection{Ideal fluids}
\label{idealf}

The relativistic hydrodynamics equations are just the conservation of
energy and momentum
\begin{equation}
\nabla_\mu T^{\mu \nu} = 0.
\label{releom}
\end{equation}	
An ideal relativistic fluid has a stress tensor given by\footnote{We
use the subscript ``rel'' for quantities in relativistic hydrodynamics
and indicate quantities in non-relativistic hydrodynamics without
subscripts.}
\begin{equation}
T^{\mu \nu} = (\er + \pr ) \, \u^\mu \, \u^\nu + \pr \, \eta^{\mu \nu} \ ,
  \label{idealtensor}
\end{equation}
where the energy density $\er$ is related to the pressure $\pr$ by a
thermodynamic equation of state.  Equations~\req{releom} and
\req{idealtensor} define a system of $d+2$ equations for the $d+2$
unknowns.

Non-relativistic ideal hydrodynamics is described by the continuity equation,
\begin{equation}
\partial_t \rho + \partial_i \left( \rho\,  \vel^i\right) = 0,
\label{continuity}
\end{equation}
together with the equation of momentum conservation, (here $i = 1,\ldots,d$)
\begin{equation}
\dt(\r\, \vel^i ) + \dj \Pi^{ij} =0, \qquad
 \Pi^{ij} = \r \, \vel^i \, \vel^j + \delta^{ij} \pnr\,,
 \label{NS}
\end{equation}
and the equation of energy conservation,
\begin{equation}
  \dt \left( \enr + \frac{1}{2}\, \rho \, \vsq \right) + \di \, \ef^i =0,
  \qquad
\ef^i =\frac{1}{2}\, ( \enr+ \pnr )\, \vsq\, \vel^i \ .
\end{equation}
where $\vsq = \vel^i\, \vel^i$.

Consider the relativistic equations \req{releom} on the light-cone.
We will consider only solutions to the relativistic equations that do
not depend on $x^-$; that is, all derivatives $\partial_-$ vanish.
The coordinate $x^+$ corresponds to the non-relativistic time $t$.
The equations of energy-momentum conservation are,
\begin{equation}
 \d_+ T^{++} + \d_i T^{+i} =0 \ , \qquad 
 \d_+ T^{+i} + \d_j T^{ij} = 0 \ , \qquad 
 \d_+ T^{+-} + \d_i T^{-i} =  0,
\label{releqns}
\end{equation}	
which reduce to the non-relativistic equations under the following
identification: identify $T^{++}$ with the mass density, $T^{+i}$ with
the mass flux (which is equal to the momentum density), $T^{ij}$ with
the stress tensor, $T^{+-}$ with the energy density, and $T^{-i}$ with
the energy flux,
\begin{eqnarray}
  && T^{++} =\rho, \quad T^{+i} = \rho \, \vel^i, \quad T^{ij} =\Pi^{ij} ,
  \nonumber \\
  && T^{+-} = \enr + \frac{1}{2}\, \rho \, \vsq, \qquad T^{-i} = \ef^i .
  \label{stiden}
\end{eqnarray}

It is now easy to convince oneself based on \req{stiden} that the
precise mapping between relativistic and non-relativistic hydrodynamic
variables is
\begin{eqnarray}
  \u^+ & = &\sqrt{ \frac{1}{2} \, \frac{\rho}{ \enr + \pnr} } \ , \qquad 
  \u^i  = \u^+ \,\vel^i ,\nonumber\\
  \pr &=& \pnr \, , \qquad \qquad\quad\;\;
  \er = 2\,\enr + \pnr.
\label{idealmap}
\end{eqnarray}
The component of the relativistic velocity $\u^-$ can be determined
using the normalization condition $\u_\mu\,\u^\mu = -1$ to be
\begin{equation}
 \u^- = \frac{1}{2}\, \left( \frac{1}{\u^+} + \u^+ \,\vsq \right) .
\label{ummap}
\end{equation}	

While the analysis has been for a general relativistic fluid with an
equation of state $\er(\pr)$, we will soon focus on conformal
fluids. Conformal invariance requires that the stress tensor for the
relativistic theory be traceless, $T^\mu_\mu =0$, which gives us the
equation of state $\er = (d+1) \, \pr$. In the non-relativistic theory
we can once again use the conformal invariance to learn that $2\, \enr
= d\, P$.
 
\subsection{Viscous fluids}
\label{viscfld}

We now wish to extend our mapping of relativistic hydrodynamics into
non-relativistic hydrodynamics to first order in derivatives on both sides.
The ideal stress energy tensor \req{idealtensor} can be supplemented
with dissipative terms, which can be expanded systematically in terms
of derivatives of the velocity field and pressure. Specifically, we
have
\begin{equation}
  T^{\mu \nu} = (\er+\pr)\, \u^\mu \,\u^\nu + \eta^{\mu \nu} \pr 
  +\pi^{\mu\nu}, \\
 \label{genstress}
 \end{equation}
where $\pi^{\mu\nu}$ incorporates all the dissipative contributions.
For first order viscous hydrodynamics we have
\begin{equation}
  \pi^{\mu\nu} = -2\, \etar\, \tau^{\mu\nu} ,
  \label{diseta}
\end{equation}
where
\begin{equation}
\tau^{\mu \nu} =   \frac{1}{2}\, P^{\mu\alpha}\, P^{\nu\beta}\, 
  \left(\nabla_\alpha \u_\beta + \nabla_\beta \u_\alpha 
  -\frac2{d+1} \, \eta_{\alpha\beta}\, \nabla_\gamma \u^\gamma\right)  
\label{taudef}
\end{equation}	
is the shear tensor and we have introduced the spatial projector
$P^{\mu\nu}=\eta^{\mu\nu}+ \u^\mu \, \u^\nu$.

We will use the zeroth-order equations of motion to simplify the
viscosity term.  By using zeroth-order equations, we make an error of
second order in derivatives, which can be neglected.  Namely, we use
the ideal hydrodynamic equations in the following form,
\begin{eqnarray}
 && \u_\mu \, \nabla^\mu \er + (\er + \pr) \, \nabla_\mu  \u^\mu = 0 ,
 \nonumber \\
 && \u_\nu \, \nabla^\nu \u^\mu + \frac{\nabla^\mu_\perp\pr}{\er + \pr} = 0,
 \qquad 
\text{where}\; \;\;   \nabla^\mu_\perp \equiv P^{\mu\alpha} \nabla_\alpha \ ,
\label{relrew}
\end{eqnarray}
to rewrite the stress-energy tensor as
\begin{eqnarray}
  T^{\mu\nu} &=& (\er+\pr) \, \u^\mu \, \u^\nu + \pr \,\eta^{\mu\nu} 
  \nonumber \\
  && \qquad -\;\etar\left(
  \nabla^\mu \u^\nu + \nabla^\nu \u^\mu 
 - \frac{2}{d+1}\, P^{\mu\nu}\, \nabla_\alpha \u^\alpha   
 - \frac{(\u^\mu \,\nabla_\perp^\nu + \u^\nu\, \nabla_\perp^\mu)\pr}
     {\er+\pr}\right).
\label{strew}
\end{eqnarray}

On the non-relativistic side, we use the ideal hydrodynamic equations in
the form
\begin{eqnarray}
 && \d_t \rho + \d_i (\rho \, \vel^i) =0, \nonumber \\
 && \d_t \vel^i + \vel^j \, \d_j\vel^i + \frac{1}{\rho}\, \d_i\pnr = 0 ,
   \nonumber \\
 && \d_t \,\enr + \d_i(\enr \, \vel^i) + \pnr \, \d_j\vel^j = 0.
\label{nr0ord}
\end{eqnarray}
The first-order contributions to the (spatial) stress tensor and the
energy flux are
\begin{eqnarray}
 & \Pi^{ij} = \rho \,\vel^i \,\vel^j + \pnr \,\delta^{ij} 
  - \eta\, \sigma^{ij}, \qquad 
   \sigma^{ij} = 
  \d_i \vel_j + \d_j \vel_i 
  - \frac{2}{d}\, \delta^{ij} \, \d_k \vel^k, \nonumber \\
 & \ef^i = \left( \enr + \pnr + \frac{1}{2}\,\rho \, \vsq \right) \, \vel^i 
  + \eta \, \sigma^{ij}\,  \vel^j -\kappa \, \d_i T,
\label{nr1ord}
\end{eqnarray}
where $\kappa$ is the thermal conductivity and $T$ is the temperature.

By using \req{strew} and \req{nr1ord}, we now establish the mapping
between relativistic and non-relativistic viscous hydrodynamics.
First, we find $\tau^{++}=0$, and therefore
\begin{equation}
  T^{++} = (\er+\pr) \,(\u^+)^2.
\end{equation}
The identification $T^{++}=\rho$ implies then that 
\begin{equation}
 \u^+ = \sqrt{\frac\rho{\er+\pr}}\,,
\label{upmapping1}
\end{equation}
unchanged from the ideal hydrodynamic level \req{idealmap}.

Next, we find
\begin{equation}
  \tau^{i+} = - \etar \( \di \u^+ - \frac{ \u^+ \, \di \pr}{2\,(\er+\pr)} \).
  \label{taupi}
\end{equation}
On the other hand, we still want to map $T^{+i}=\rho \,\vel^i$.  This
means that there is now a correction to the relation between $\u^i$
and $\vel^i$:
\begin{equation} \label{upmapping2}
  \u^i = \u^+ \left[ \vel^i + \frac{\etar}{\rho}\, \left( \d_i \u^+ - 
        \frac{\u^+ }{2\,(\er+\pr)}\,\d_i\pr\right)\right].
\end{equation}
For $T^{ij}$, after some algebra, we find
\begin{equation}
   T^{ij} = \rho\, \vel^i \,\vel^j + \pr\, \delta^{ij} 
   -\etar \, \u^+ \left(\d_i \vel_j +\d_j \vel_i 
   - \frac{2}{d} \, \delta^{ij} \, \d_k \vel^k\right),
   \label{Tij}
\end{equation}
which implies that the pressures on the two sides still coincide,
\begin{equation}
  \pr = \pnr,
  \label{presmap}
\end{equation}
and the relationship between the viscosities is
\begin{equation}
  \etar = \frac{\eta}{\u^+}.
  \label{etamap}
\end{equation}
Note that our identifications automatically give a first-order
correction $\sigma_{ij}$ in the non-relativistic theory with the
correct tensor structure. That is, the trace-free relativistic shear
tensor gives a trace free spatial stress tensor in the
non-relativistic theory.

Regarding the other components of the stress tensor, after some
calculations involving many cancellations, one discovers that $T^{+-}$
is
\begin{equation}
  T^{+-} = \frac{1}{2}\, (\er-\pr) + \frac{1}{2}\, \rho\,  \vsq,
\end{equation}
which means that the relationship between relativistic and
non-relativistic energy densities remain unchanged,
\begin{equation}
  \er = 2\,\enr + \pnr.
\end{equation}
Finally, for $T^{-i}$ we find
\begin{equation}
  T^{-i} = \left( \enr + P + \frac{1}{2}\, \rho \, \vsq\right) \, \vel^i   
  - \etar \,\u^+ \sigma^{ij} \,\vel_j  
  + \etar\, \delta^{ij}\, \frac{\d_j \u^+}{(\u^+)^2}
  - \etar \, \frac{\u^+}{\rho}\,\delta^{ij}\, \d_j P.
\end{equation}
Thus we have to require that 
\begin{equation}
  \etar \,\frac{\d_i \u^+}{(\u^+)^2} - \etar \,\frac{\u^+}{\rho}\, \d_i P
  = -\kappa \,\d_i T.
  \label{kappamapping1}
\end{equation}
In order to see that the left hand side is indeed proportional to
$\d_i T$, we need to use the mapping \req{upmapping1} and the
equation of state for a non-relativistic theory.  

Focusing specifically now on conformally invariant fluids, using
\req{upmapping1} and $\enr=\frac{d}{2}\, P$, we find
\begin{equation}\label{lhs-kappa}
  \etar \,\frac{\d_i \u^+}{(\u^+)^2} - \etar \,\frac{\u^+}{\rho}\, \d_i P = 
  - \etar \,\sqrt{\frac{\enr+P}{2\, \rho}}\; 
  \d_i \ln \left(\frac{P^{(d+4)/(d+2)}}{\rho}\right).
\end{equation}
Recalling that the equation of state of the holographic
non-relativistic liquid is \citep{Kovtun:2008qy}
\begin{equation}
  P = \alpha \left(\frac{T^2}{\mu}\right)^{(d+2)/2},
\end{equation}
the argument of the logarithm in \req{lhs-kappa} is $T^2$ up to a
constant.  Therefore, the left hand side of \req{kappamapping1} is
indeed proportional to $\d_i T$, and one reads out the value for the
thermal conductivity:
\begin{equation}
\kappa = 2\,\eta\, \frac{\enr+P}{\rho\,  T}.
\label{kappaval}
\end{equation}

Let us now compute the Prandtl number.  The Prandtl number is defined
as the ratio of the kinematic viscosity $\nu$ and the thermal
diffusivity $\chi$,
\begin{equation}
  {\rm Pr} = \frac\nu\chi,
 \label{Prdef}
\end{equation}
where
\begin{equation}
  \nu = \frac{\eta}{\rho}\ , \qquad \chi = \frac{\kappa}{\rho \, c_p},
\end{equation} 
where $c_p$ is the specific heat at constant   pressure.  We note 
the definition of the heat capacity at constant volume:
\begin{equation}
  C_v = \left(\frac{\d H}{\d T}\right)_{P,N},
\end{equation}
where $H=E+PV$ is the enthalpy. Write $H=w\, V=w\,N/n$. We then find
\begin{equation}
  C_p = N\, \frac\d{\d T} \left( \frac w n \right)_P
      = -\frac{N\, w}{n^2} \left(\frac{\d n}{\d T}\right)_P
\end{equation}
where we have used the fact that $w=(d/2+1)\, P$ and is fixed at fixed
$P$.  At fixed $P$, $\mu\sim T^2$, and $n=\d P/\d T\sim 1/T^2$, and
$\d n/d T=-2 n/T$.  Therefore
\begin{equation}
  C_p = \frac{2\,w\, N}{T\,n}
\end{equation}
and $c_p=C_p/M$ ($M$ being the total mass) is equal to $2w/\rho\, T$.
Thus we find:
\begin{equation}
  {\rm Pr} = \frac{2\,w \,\eta}{\rho\, T\,\kappa} =1.
\end{equation}
Note that this result is valid for any non-relativistic conformal
fluid obtained from the DLCQ of a relativistic conformal fluid.

\section{Thermal description of non-relativistic CFTs}
\label{gravdisc}

Non-relativistic conformal field theories with Schr\"odinger symmetry
in $d$ spatial dimensions are dual to $\Schr{d+3}$ spacetime
\citep{Son:2008ye,Balasubramanian:2008dm}. The $\Schr{5}$ spacetimes
can be realized in string theory as the near-horizon geometry of
D3-branes probing a Null Melvin universe
\citep{Herzog:2008wg}.\footnote{The near-horizon limit in this case
needs to be taken keeping in mind that we want to realize the scaling
symmetry consistent with the Schr\"odinger algebra. We thank James
Lucietti for a useful discussion about this issue.} Furthermore, this
background can be obtained as a solution to a 5-dimensional Lagrangian
which is a consistent truncation of IIB supergravity
\citep{Maldacena:2008wh}, involving gravity coupled to a massive vector
field and three scalars:
\begin{eqnarray}
  \CS_{\rm bulk}  &=& \frac{1}{16 \, \pi G_5}\, \int \, d^5x\, \sqrt{-g} \, 
  \left( R + V(\phi_i) - 5 \, (\partial \phi_1)^2 
  - \frac{15}{2}\, (\partial \phi_2)^2 
  - \frac{1}{2}\, (\partial \phi_3)^2 \right.\nonumber \\
&& \qquad \left.
 - \frac{1}{4}\, g(\phi_i) \, F_{\mu\nu}F^{\mu \nu} - 
4 \, e^{-2\phi_1 -3 \phi_2 -\phi_3} \, A_\mu A^\mu \right), \nonumber \\
  V(\phi_i) &=& 24 \, e^{-\phi_1 - 4 \,\phi_2} - 4\, e^{-6\,\phi_1 
    - 4 \,\phi_2} - 8\, e^{-10\,\phi_2}, \nonumber \\
  g(\phi_i) &=&e^{4\phi_1 + \phi_2 -\phi_3}.
\label{mmtact}
\end{eqnarray}	
In fact, all known solutions are solutions to a slightly simpler theory
outlined in \citep{Herzog:2008wg}, where the three scalars are linearly
related as
\begin{equation}
\{\phi_1 , \phi_2, \phi_3\} = \{ -\frac{2}{5}, -\frac{1}{15}, 1\} \, \phi.
\label{scalarrel}
\end{equation}	
This allows one to consider the five-dimensional effective action
\begin{equation}
  S = \frac{1}{16 \, \pi G_5}\,\int d^5 x \sqrt{-g} \left(R 
  - \frac{4}{3} (\partial_\mu \phi) (\partial^\mu \phi) 
  - \frac{1}{4} e^{-8 \phi / 3} F_{\mu\nu} F^{\mu\nu} 
- 4 A_\mu A^\mu - V(\phi) \right)  ,
\label{5deff}
\end{equation}
where
\begin{equation}
V(\phi) = 4 \,e^{2 \phi/3} (e^{2 \phi} - 4).
\label{Vpot}
\end{equation}	

The field theory on the D-branes in the appropriate decoupling limit
is $\CN =4$ Super Yang-Mills plus a non-commutative
deformation, giving rise to a version of the dipole field theories discussed in
\citep{Bergman:2000cw,Alishahiha:2003ru}. Consider the fields in $\CN
=4$ SYM (with gauge group $SU(N)$) which transform under the global
symmetry $SO(4,2) \times SO(6)$. Picking a $U(1) \subset SO(6)$ we
deform the field theory by replacing ordinary products appearing in
the Lagrangian by a star-product \cite{Maldacena:2008wh}
\begin{equation}
f\star g = e^{i \,\beta\, \left(P_-^f \, R^g - P_-^g \, R^f\right)} \, f g
\label{starprod}
\end{equation}	
where $P_-$ is the momentum along the null $x^-$ direction and $R$ is
the $U(1)$ charge. This product is a hybrid between the usual
non-commutative star-product which only involves spatial momenta and
the $\beta$-deformation of $\CN=4$ which involves only the
R-charges. Of importance later will be the fact that this deformed
field theory inherits some of the properties from $\CN =4$ SYM. In the
large $N$ limit we in fact expect that the planar sector of the
deformed theory to be identical to that of $\CN =4$ SYM \cite{Maldacena:2008wh}.\footnote{As discussed in \sec{intro} these statements extend to generic $\CN =1$ superconformal field theories which we deform by a $U(1)_R$ symmetry.}

The Lagrangian \req{5deff} has a black hole solution which was
obtained by the Null Melvin Twist \citep{Herzog:2008wg, Adams:2008wt}
or TsT solution-generating transformation \citep{Maldacena:2008wh} in
10 dimensions.  The black hole geometry is given as
\begin{eqnarray}
 ds_E^2 &=&   r^2\, k(r)^{-\frac{2}{3}}\left( \left[\frac{1-f(r)}{4\,\beta^2} -
    r^2\,f(r)\right] \, (dx^+)^2 
  + \frac{\q }{r^4} \, (dx^-)^2 
  - \left[1+f(r)\right]\,dx^+\,dx^- \right) \nonumber \\
  && \quad+\;\;  k(r)^{\frac{1}{3}}\, \left(r^2 d {\bf x}^2 
  +  \frac{dr^2}{r^2\, f(r)} \right),
\label{5dbh}
\end{eqnarray}
with the massive vector and scalars taking the form
\begin{eqnarray}
  A &=& \frac{r^2 }{k(r)} \, \left( \frac{1+f(r)}{2}\, dx^+
  - \frac{\q }{r^4}\, dx^- \right), \nonumber \\
  e^\phi &=& \frac{1}{\sqrt{k(r)}} \ ,
\label{5doth}
\end{eqnarray}
where $f(r)$ and $k(r)$ are 
\begin{equation}
f(r) = 1- \frac{r_+^4}{r^4} \ , \qquad k(r) = 1+ \frac{\q}{r^2} \, ,
\label{fkdefs}
\end{equation}	
with $\q \equiv \beta^2 \, r_+^4$. Note that in these light-cone
coordinates, the solution asymptotically approaches the vacuum
solution \req{vacmet} at large $r$, and also reduces to \req{vacmet} when we set $r_+=0$. 
It has been argued in
\citep{Herzog:2008wg} that the black hole spacetime \req{5dbh}
corresponds in the field theory to a grand canonical ensemble at
temperature
\begin{equation}
  T = \frac{r_+}{\pi \,\beta}
\label{bhtemp}
\end{equation}	
and a chemical potential for particle number
\begin{equation}
\mu = \frac{1}{2 \,\beta^2} \ .
\label{mudef}
\end{equation}	
Note that the Schr\"odinger algebra involves a conserved charge
associated with particle number, which geometrically is realized via
the Killing field $\partial_-$. These results were obtained with minor differences in the derivation in \cite{Maldacena:2008wh,Adams:2008wt}.

Using a Euclidean action calculation, the authors of
\citep{Herzog:2008wg} derived the conserved charges of the black
hole. This calculation required a careful analysis of the boundary counter-terms, since the metric \req{5dbh} has rather complicated
asymptotics. In \App{appHam} we discuss the calculation of the
conserved charges in a Hamiltonian formulation, supplementing the
analysis of \citep{Herzog:2008wg}. The conserved charges associated
with the Killing symmetries $\partial_-$ and $\partial_+$ in
\req{5dbh} translate in the field theory to particle number $N$ and
total energy $E$. To obtain finite values we assume that the
$x^-$-direction is compactified with period $\Delta x^-$.  The results
obtained in \citep{Herzog:2008wg} are
\begin{equation}
\langle N \rangle =  \langle P_- \rangle \frac{\Delta x^-}{2\pi}=\frac{\pi^2 \,
  T^4}{64\,G_5\, \mu^3} \;V\, (\Delta x^-)^2 \ = \frac{\gamma^2}{8 \pi^2 \,G_5} \;V\, (\Delta x^-)^2 ,
\label{partno}
\end{equation}
and
\begin{equation}
\langle E \rangle = \frac{\pi^3 \,T^4}{64\, G_5\,\mu^2} \; V\, \Delta x^- \ = \frac{r_+^4}{16 \pi \,G_5} \; V\, \Delta x^- ,
\label{energy}
\end{equation}
where we have given the results in terms of the physical temperature
and chemical potential, and also in terms of the parameters in the
bulk solution to emphasize the physical interpretation of the
parameters $\gamma$ and $r_+$. Furthermore, the pressure is given in
the grand canonical ensemble directly in terms of the Gibbs potential
$\CQ(T,\mu,V)$:
\begin{equation}
P\, V = - \CQ(T,\mu, V) =  \frac{\pi^3 \,T^4}{64\, G_5\,\mu^2} \; V\, 
\Delta x^- \ ,\label{pressure}
\end{equation}	
leading thus to an equation of state
\begin{equation}
P \, V =  E \;\; \thus \;\;  P = \enr.
\label{eos}
\end{equation}	
In addition, we note for future reference that the entropy of the black
hole is given by
\begin{equation}
S =  \frac{\pi^3 \,T^3}{16 \,G_5\,\mu^2} \,V\, \Delta  x^-\ .
\end{equation}
%

\section{Hydrodynamic description of non-relativistic CFTs}
\label{nrfluids}
The black hole solution \req{5dbh} corresponds to an equilibrium
configuration of the non-relativistic field theory. We would like to
study departures from equilibrium in the continuum limit by
considering local patches of the field theory in local
equilibrium. This is the hydrodynamic limit, where one has local
variations of energy density and particle number; these local domains
evolve according to the laws of hydrodynamics which were described in
\sec{lcredn}.

To obtain a gravitational description of the fluid dynamical regime,
we need to patch together local domains of equilibrated fluid -- the
precise manner in which this can be achieved was outlined in
\citep{Bhattacharyya:2008jc}. The authors presented an algorithmic
procedure to construct gravitational solutions starting from the
equilibrium black hole solution. The idea is to consider the most
general stationary solution for the equilibrium -- for the case of
relativistic superconformal theories in $d+2$ dimensions, this is
given by a boosted Schwarzschild-$\AdS{d+3}$ black hole, which is
specified by $d+1$ parameters: a horizon size $r_+$ and a unit
normalized velocity field $\u_\mu$. Choosing Eddington-Finkelstein
like ingoing coordinates which are regular on the horizon, the metric
takes the form:
\begin{equation}
ds^2 =2 \,\u_\mu \,dx^\mu dr - r^2\, f(r)\, \u_\mu \u_\nu \;dx^\mu dx^\nu + r^2
\,P_{\mu\nu} \,dx^\mu dx^\nu 
\label{sads}
\end{equation}
where $P_{\mu \nu} = \eta_{\mu \nu} + \u_\mu \, \u_\nu$ is the spatial
projector introduced earlier and $f(r) = 1 -(r_+/r)^{d+2}$.

To study the hydrodynamic configurations one promotes the parameters
$r_+$ and $\u_\mu$ to functions of the boundary coordinates
$x^\mu$. Then one recursively solves for the bulk metric order by
order in a boundary derivative approximation. This procedure was
systematically carried out to second order in
\citep{Bhattacharyya:2008jc}. To ensure regularity of the bulk
solution, and in particular to guarantee the existence of a regular
future event horizon, it was necessary to adapt coordinates wherein
lines of constant boundary $x^\mu$ corresponded to ingoing null
geodesics in the bulk. This coordinate chart was utilized in
identifying how the locally equilibrated boundary domains evolve in
the bulk radial direction -- the fluid in such a domain was evolved
along a tube in the bulk centered about the ingoing null
geodesic. The size of these tubular domains is set by the local
energy density and they provide the bulk analog of patching together
pieces of equilibrated fluid.

\subsection{Gravitational dual of a non-relativistic fluid: 
Direct construction}

We now turn to the analogous calculation for the non-relativistic CFT
discussed in \sec{gravdisc}. We should first write the metric
\req{5dbh} in a form that is regular through the future horizon. To do
so, perform a coordinate transformation
\begin{equation}
x^+ \to x^+ + \beta \, p(r) \ , \qquad x^- \to x^- +
\frac{1}{2\,\beta} \, p(r)
\end{equation}	
with $p(r)$ chosen such that the metric is regular on the horizon $r =
r_+$. A convenient choice is\footnote{If one wanted further
$\partial_r$ to be null then one could instead choose $p'(r) = \pm
\sqrt{\frac{k(r)}{1+ \beta^2 \, r^2} }\,\frac{1}{r^2\, f(r)}$.}
\begin{equation}
p'(r) = \frac{1}{r^2\, f(r)}\,,
\end{equation}	
leading to a metric
\begin{eqnarray}
ds^2 &=& r^2\, k(r)^{-\frac{2}{3}}\left(\left[\frac{1-f(r)}{4\beta^2} -
    r^2\,f(r)\right] \, (dx^+)^2 + \frac{\q}{r^4} \, (dx^-)^2 
    - \left[1+f(r)\right]\,dx^+\,dx^- \right) \nonumber \\
&& -  k^{-\frac{2}{3}}\, 
  \left[\left(\frac{1}{\beta}+2\,\beta\, r^2\right) \, dx^+ 
  + 2\, \beta \,dx^- \right]\, dr -  k^{-\frac{2}{3}}\, \beta^2\, dr^2
  + k(r)^{\frac{1}{3}}\, r^2 d {\bf x}^2 \ .
\label{5dbhed}
\end{eqnarray}	

This gives a two-parameter family of solutions of the 5d effective
Lagrangian \req{5deff}. We can construct a four parameter family of
solutions trivially by performing a boost via the coordinate
transformation
\begin{equation}
{\bf x} \to {\bf x} +  {\bf \vel}\, x^+ \ , \qquad 
  x^- \to x^- + {\bf \vel}\cdot {\bf x} + \frac{1}{2}\, \vsq \,x^+.
\label{galboost}
\end{equation}	
Since the background metric \req{vacmet} has Galilean invariance, this
boost does not change the asymptotic form of the metric.

This boosted black hole, characterized by the parameters $\{r_+,
\beta, {\bf \vel}\}$, is the starting point for a systematic hydrodynamic
analysis. Following \citep{Bhattacharyya:2008xc} we should promote the
parameters $\{r_+, \beta, {\bf \vel}\}$ to fields depending on $(x^+, {\bf x})$
and consistently solve the equations of motion order by order in
derivatives in the $x^+$ and ${\bf x}$ directions. This process, while
straightforward, is rendered cumbersome by the presence of additional
fields in the action \req{mmtact}. We will therefore resort to a trick
to recover the dual geometry.

\subsection{Inhomogeneous black holes via TsT transform}
\label{nmthydro}


The solution we want to obtain is characterized by four parameters
depending on the coordinates $(x^+, {\bf{x}})$. The general
inhomogeneous black hole of \citep{Bhattacharyya:2008jc} also depends
on four parameters, $r_+$ and $\u_\mu$, and if we specialize to
solutions independent of $x^-$, they also depend on the coordinates
$(x^+, {\bf{x}})$. For such solutions, we can then apply the same TsT
transformation used to obtain the black hole solution \req{5dbh} to
obtain inhomogeneous black holes with Schr\"odinger asymptotics. Since
these solutions will by construction reduce to \req{5dbh} when the
parameters are constants, we argue that they are precisely the
required inhomogeneous solutions. Below, we discuss the action of the
TsT transformation on the solution of \citep{Bhattacharyya:2008jc}. 

The TsT transformation is a solution-generating technique in string
theory wherein one uses a twisted T-duality to add NS-NS flux to a
given solution \citep{Lunin:2005jy}.  Consider a solution to type IIB
supergravity of the form $\CM_5 \times \CX_5$ where $\CX_5$ is a
Sasaki-Einstein space, which we view as a $U(1)$ fibration over a base
$\CB $, and $\CM_5$ is a constant negatively curved spacetime, which we
will take to be asymptotically $\AdS{5}$. The simplest case is when
$\CX_5 = \Sp^5$ and the base is thus a ${\bf CP}^2$. We assume that $\CM_5$
admits a Killing vector field $\partial_-$, so that $\CM_5 \times
\CX_5$ has an isometry group $U(1)_{x^-} \times U(1)_\psi$. Viewed as
an eight dimensional solution upon Kaluza-Klein reduction, we have an
$SL(2,\R)$ symmetry group, which can be used to generate new
solutions. This is the TsT transformation. 

We start from the solutions dual to arbitrary fluid flow for
hydrodynamics of relativistic superconformal theories of
\citep{Bhattacharyya:2008jc},
\begin{equation}
ds^2 = g_{AB}\, dX^A \, dX^B = -2 \, \u_\mu\, \CS(r,x)\, dx^\mu \,dr 
  + \chi_{\mu\nu}(r,x)\, dx^\nu\, dx^\nu ,
\label{flumet}
\end{equation}	
where $X^A=\{x^\mu, r\}$. The metric dual to viscous fluid
dynamics for the relativistic conformal fluids is given by
\req{flumet} with $\CS(r,x)=1$ and 
\begin{equation}
\chi_{\mu \nu} =  r^2\,\left(P_{\mu \nu} 
-  f(r,x)\, \u_{\mu}\, \u_{\nu}\right) 
+ \frac{2 }{r_+} \, r^2\,F(r,x)\, \tau_{\mu\nu}  
+ \frac23 \, r \, \u_{\mu}\, \u_{\nu} \,\nabla_{\lambda} \u^{\lambda}  
-  r\, \u^{\lambda}\nabla_{\lambda}\left(\u_\nu \u_{\mu}\right) ,
\label{replmet}
\end{equation}
where $P_{\mu \nu}$ is the spatial projector defined after
\req{taudef}, $f(r,x)$ is the function $f(r)$ appearing in \req{fkdefs}
with $r_+ \to r_+(x)$, and 
\begin{equation}
\label{fdef}
F(r,x)  =\frac14\,
\left[\ln\left(\frac{(r+r_+(x))^2(r^2+r_+(x)^2)}{r^4}\right) -
  2\,\arctan(r/r_+(x)) +\pi\right] . 
\end{equation}	
It is important to note that $\u_\mu$ and $r_+$ are no longer
parameters, but regarded as functions of $x^\mu$.  To apply the TsT
transformation, we need to choose a direction $x^-$, and assume $r_+,
\u_\mu$ are independent of $x^-$.

The general map between the parameters of a relativistic fluid and the
parameters of a non-relativistic fluid to first order in derivatives
was written in \req{upmapping1}, \req{upmapping2} and
\req{presmap}. When we write the relativistic fluid in terms of the
gravitational dual spacetime, we can rewrite this mapping as a mapping
between the functions $r_+, \u_\mu$ characterizing the asymptotically
AdS spacetime and the functions $r_+, \beta, \vel_i$ appropriate in
the non-relativistic case.  The function $r_+$ is the same in both
descriptions, and the map from $\u^\mu$ to $\beta$, $\vel^i$ is
\begin{equation} \label{bulkmapp}
\u^+ = \beta, \quad \u^i =  \beta \left[ \vel^i +
  \frac{1}{4 \beta^2} \partial_i \frac{\beta}{r_+} \right]. 
\end{equation}

The full 10 dimensional metric is a direct sum of the metric $g_{AB}$
on $\CM_5$ given in \req{flumet}  and a Sasaki-Einstein space $\CX_5$,
\begin{eqnarray}
ds_E^2 &=& g_{AB}\, dX^A\, dX^B +h^2\, \left(d\psi + A\right)^2 + ds^2(\CB),
\nonumber \\
F_{(5)} &=& 4\, \left({\rm Vol}(\CM_5) +h\, {\rm Vol}(\CB) \wedge (d\psi + A) 
\right).
\label{10dstart}
\end{eqnarray}	
Under the TsT transformation the metric \req{10dstart} gets mapped to
a new solution of Type IIB supergravity
\citep{Maldacena:2008wh}\footnote{In \citep{Maldacena:2008wh}, the TsT
  transformation involved an arbitrary parameter $\sigma$; however,
  this can be absorbed into a redefinition of the coordinate $x^-$ by
  a boost in the $x^\pm$ plane. We have found it more transparent to
  fix the parameter in the TsT transformation and keep instead the
  velocity $\beta$ in \req{bulkmapp} as the free parameter
  corresponding to the choice of $x^-$.}
\begin{eqnarray}
ds_E^2 &=& e^{-\frac{\varphi}{2}} \, \left( g_{AB}\, dX^A\, dX^B  
-\, e^{2\,\varphi}\, h^2 \, g_{A-}\,g_{B-}\, dX^A dX^B
+  e^{2\,\varphi}\,h^2\,\left(d\psi + A\right)^2 + ds^2(\CB)\right), \nonumber \\
F_{(5)} &=& 4\, \left({\rm Vol}(\CM_5) 
+ {\rm Vol}(\CB) \wedge h \,(d\psi + A) \right), \nonumber \\
B_{(2)} &=&  e^{2\,\varphi}  \,h^2\,  g_{B-}\, dX^B \wedge (d\psi + A) ,
\nonumber \\
e^{-2\, \varphi} &=& 1 +h^2\,  g_{--} \ .
\label{tstedsol}
\end{eqnarray}	
This solution can be Kaluza-Klein reduced back to five dimensions to
give a metric (restricting to situations where the norm of the Reeb vector $\partial_\psi$ is fixed to $h^2 =1$),
\begin{equation}
ds_5^2 = e^{-\frac{2}{3}\, \phi} \, \left(g_{AB} -  e^{2\, \phi} \, 
g_{A-}\, g_{B-}\,\right)\, dX^A\, dX^B ,
\end{equation}	
supported by 
\begin{eqnarray}
A &=&  e^{2\phi}\, g_{A-}\, dX^A, \nonumber \\
e^{2\,\phi} &=& \frac{1}{1+ g_{--}}.
\end{eqnarray}	
As before this five dimensional solution solves the equations of
motion arising from \req{mmtact} with the scalars being related as in
\req{scalarrel}.

Restricting to configurations which have
$\partial_-$ as an isometry, after a TsT transformation on the metric \req{flumet}
we get a new metric of the form 
\begin{eqnarray}
  ds_E^2 &=& e^{-\frac{2}{3} \, \phi} \, \left( 
  -2\, \u_\mu\, \CS\, dx^\mu \, dr + \left[\chi_{AB} 
  - \frac{{\widetilde\chi}_A\, {\widetilde \chi}_B}
  {1+ \chi_{--}}  \right]\, dX^A\, dX^B \right), \nonumber \\
    A &=&  e^{2\phi}\, \widetilde{\chi}_A\, dX^A, \nonumber \\
  e^{2\,\phi} &=& \frac{1}{1+ \chi_{--}} ,
\label{nrflumet}
\end{eqnarray}	
with 
\begin{equation}
{\widetilde \chi}_A = \chi_{A-} - \u_- \, \CS \, \delta_A^r \ .
\label{tchidef}
\end{equation}	
The TsT transform converts the asymptotically $\AdS{5}$ spacetime
\req{flumet} to an asymptotically $\Schr{5}$ spacetime, which depends
on the $r_+, \beta, \vel_i$ defined in \req{bulkmapp} which are
arbitrary functions of $(x^+, {\bf x})$. This provides the required
inhomogeneous generalization of the black hole solution \req{5dbh}. We
will not write the result of applying the transformation more
explicitly, as it is quite complicated, and its construction is a
straightforward exercise.  Also note that since the internal manifold $\CX_5$ plays a minimal role in our construction, it is easy to verify that the solution \req{nrflumet} also solves the consistent truncation action \req{mmtact} with the scalars still related according to \req{scalarrel}.

\subsection{Properties of black holes dual to non-relativistic fluids}
\label{nrbhprop}

We will now discuss some of the physical properties of the solution
\req{nrflumet}. The solutions \req{nrflumet} are the most general
long-wavelength regular solutions dual to configurations of the dual
non-relativisitic conformal field theory and are valid to leading order in the boundary derivative expansion. These geometries solve the field equations arising  from \req{5deff} provided the boundary stress tensor complex satisfies the non-relativistic Navier-Stokes equations. This follows from the fact
that \req{flumet} are the most general regular solutions dual to the
relativistic field theory and that the regularity properties of the
black hole solutions are unaffected by the TsT transformation. The
regularity of the solutions \req{flumet}, in particular, the fact that
they have a regular event horizon, was demonstrated in
\citep{Bhattacharyya:2008xc}. This result required that the variations
in the boundary directions parameterized by $x^\mu$ are slow. In
employing the TsT transformation, all we required was that
$\partial_-$ be a Killing vector; so there is no variation of the
relativistic fluid in the light cone direction and one can of course
take the variations in all the other directions to be appropriately
slow.  To explicitly demonstrate that the solution \req{nrflumet} has
a regular event horizon, one can follow the perturbative construction
of \citep{Bhattacharyya:2008xc}. From this analysis it is easy to infer
that for viscous non-relativistic fluids, the location of the horizon
will remain at $r=r_+(x^+,{\bf x})$.

In analyzing the regularity of the geometries constructed in
\citep{Bhattacharyya:2008jc}, it was important to work in a well
behaved coordinate chart. As explained there and subsequently
elaborated in \citep{Bhattacharyya:2008xc,Bhattacharyya:2008mz} the
solutions can be thought of as being tubewise approximated by a
homogeneous black hole solution, with the tubes being domains in the
bulk centered around radially ingoing null geodesics with width set by
the scale of variation in the boundary directions $x^\mu$. In the
non-relativistic case \req{nrflumet}, the metric is not \textit{a priori}
written in coordinates adapted to the radially ingoing geodesics. To
see this note that the gauge choice employed in
\citep{Bhattacharyya:2008jc, Bhattacharyya:2008xc} was to set $g_{r\mu}
\propto \u_\mu$.\footnote{In fact, as discussed in
\citep{Bhattacharyya:2008mz}, one can simplify the metric further, by
demanding that the radially ingoing null geodesics are affinely
parameterized by $r$. This is equivalent to setting $\CS(r,x) =1$ in
\req{flumet}.} On the other hand, in the non-relativistic case we have
non-trivial $g_{rr}$ arising after the TsT transformation coming from
the non-vanishing $\widetilde{\chi}_r$. This can of course be removed
by a coordinate transformation, and then the issue of regularity boils
down to the analysis presented in \citep{Bhattacharyya:2008xc}.

To complete the fluid-gravity correspondence, we need to identify the
stress tensor complex associated with the inhomogeneous solutions
constructed above.  The boundary field theory dual to the $\Schr{5}$
background was argued to be a non-commutative deformation of $\CN =4$
SYM. This fact allows us to argue that the planar sector of the
deformed theory with non-relativistic invariance is in fact identical
to the parent $\CN =4$ theory \citep{Maldacena:2008wh}, and the direct
computations of the thermodynamics in \citep{Herzog:2008wg} are
consistent with this picture. We therefore argue that, as advocated in
\citep{Maldacena:2008wh}, we can identify the non-relativistic stress
tensor complex corresponding to the geometry after the TsT
transformation with the one we have prior to the transformation. That
is, the dual stress tensor complex is obtained simply by performing
the DLCQ reduction described in \sec{lcredn} for the
relativistic stress tensor corresponding to the geometry \req{flumet}
before the TsT transformation. 

More explicitly, the relativistic stress tensor corresponding to
\req{flumet} (to first order in derivatives) is of the form
\req{genstress} with\footnote{Note that $\frac{1}{16 \pi \, G_5}$ gives the effective central charge of the dual field theory. For the case of deformed $\CN =4$ SYM to the non-relativistic dipole theory, this evaluates to $\frac{N^2}{8\,\pi^2}$ where $N$ is the rank of the gauge group.} 
\begin{equation}
  \er =3\, \pr, \quad \pr = \frac{r_+^4}{16 \pi \, G_5}, \quad \etar =
  \frac{r_+^3}{16 \pi \, G_5}.
\end{equation}
Thus, using the identifications in section \ref{lcredn}, the
non-relativistic stress tensor complex dual to the geometry
\req{nrflumet} will be of the form \req{nr1ord} with\footnote{Note
  that the stress tensor complex obtained from the relativistic stress
  tensor in \sec{lcredn} is a local density in the $x^-$
  direction. From the non-relativistic field theory point of view, it
  is more natural to multiply by $\Delta x^-$ to obtain an object
  which is a local density only in the spatial directions.}
\begin{equation}
\enr =  \pnr, \quad \pnr = \frac{r_+^4}{16 \pi \, G_5}, \quad \rho =
\frac{\beta^2 r_+^4}{4\pi \, G_5},
\quad \eta = \frac{r_+^3 \beta}{16 \pi \, G_5}, \quad \kappa =
\frac{r_+^2}{16 \,G_5}.
\end{equation}

The boundary stress tensor complex for the solution \req{nrflumet}
comprises a spatial stress tensor $\Pi_{ij}$, particle density $\rho$,
an energy flux $\ef^i$, momentum flux $\rho\, \vel^i$ and energy
$\enr$. While we have argued that in the large $N$ planar limit these
quantities can be derived by reducing the relativistic stress tensor
on the light-cone (using thus the properties of the TsT
transformation), it should in principle also be possible to obtain
these by direct computation in the geometry \req{nrflumet}. It turns
out to be easy to use the counter-term construction proposed in
\citep{Herzog:2008wg} to extract the spatial stress tensor without
trouble. However, it is not clear how to calculate the energy $\eps$
and the particle density $\rho$ in an analogous fashion. In
\citep{Herzog:2008wg}, the energy $\eps$ and the particle density
$\rho$ for the system in thermal equilibrium were obtained by a
Euclidean action calculation. In \App{appHam}, we discuss the
calculation of these quantities using canonical methods.

\section{Discussion}
\label{discuss}

We have discussed the hydrodynamic limit of $d$ spatial dimensional
non-relativistic conformal field theories and their dual gravitational
solutions, which are inhomogeneous black holes with $\Schr{d+3}$
asymptotics. We employed the fact that starting with a relativistic
hydrodynamical system one can obtain non-relativistic fluid dynamical
equations upon an appropriate light-cone reduction. Relativistic
hydrodynamics in $d+2$ spacetime dimensions descends to
non-relativistic hydrodynamics in $d$ spatial dimensions. We
demonstrated this both for ideal fluids and also for dissipative
fluids where we restricted attention to first order in the gradient
expansion.  The latter allowed us to recover the heat conductivity of
the non-relativistic system in terms of the shear viscosity and state
parameters of the parent relativistic fluid. In particular, for
non-relativistic CFTs we have shown that the Prandtl number is
unity.\footnote{For comparision,
Pr(water) $\approx 7$, Pr(mercury) $\sim
10^{-3}$, and Pr(air) $\approx0.7$.} Since the first
order dissipative coefficients for the non-relativistic CFTs arise from
the shear viscosity of the relativistic system, it is not surprising
that the rates of momentum diffusion and thermal conduction are
correlated.

Our construction of the geometries dual to the non-relativistic fluids
uses as its starting point the asymptotically AdS spacetimes dual to
relativistic hydrodynamics. Starting with the solutions constructed in
\citep{Bhattacharyya:2008jc} we were able to construct inhomogeneous
black holes with Schr asymptotics using the TsT transformation
described in \citep{Maldacena:2008wh}. The TsT transform deforms the
boundary field theory to a non-local quantum field theory through the
introduction of a star-product \req{starprod} -- we are therefore
looking at the hydrodynamic limit of these deformed superconformal
field theories.

Given these black hole solutions we can in principle try to extract
the non-relativistic stress tensor complex of the dual field theory
using an appropriate boundary stress tensor construction. Instead of
doing so, we adopted the philosophy advocated in
\citep{Maldacena:2008wh} -- the hydrodynamics of the non-relativistic
theory is given by the light-cone reduction of the undeformed
superconformal theory in the planar limit. This can be justified by
recalling that for superconformal field theories with gauge group of
rank $N$, the star-product deformation resulting from the TsT
transformation leaves the planar sector of the theory unchanged
\citep{Lunin:2005jy}. This allows us to efficiently extract the stress
tensor for the non-relativistic viscous fluid.  One should in
principle explicitly compute the stress-tensor complex using the
counter-term subtraction scheme discussed in \citep{Herzog:2008wg}. It
turns out to be easy to check that one can indeed extract the spatial
stress tensor. Furthermore, by rewriting the action in a Hamiltonian
formulation we have been able to extract the energy and particle
density. But we have encountered some difficulties in proving that the
Hamiltonian generates the time-translation symmetry of the
non-relativistic field theory. This is an interesting open problem
which we hope to return to in the future. It is also worth remarking
that to compute equilibrium thermodynamics we can take a simpler route
proposed in \citep{Kovtun:2008qy} -- use a background subtraction
scheme where the reference spacetime is not the vacuum $\Schr{d}$
spacetime, but corresponds to a state with zero temperature at finite
particle density.\footnote{In the notation used in
  \citep{Herzog:2008wg} this corresponds to the limit $r_+\to0$ and
  $\beta \to \infty$ with $\gamma = \beta\, r_+^2$ held fixed.}

Our discussion of the light-cone reduction of the relativistic
equations was restricted to ideal and viscous fluids, i.e., up to and
including the first order in the gradient expansion. Fluid dynamics in
general can be viewed as an effective field theory with an infinite
number of irrelevant terms obtained as usual in a derivative
expansion. It would be interesting to understand the light-cone
reduction of the second order relativistic conformal hydrodynamic
stress tensor constructed in \citep{Baier:2007ix,Bhattacharyya:2008jc}
(see \citep{Haack:2008cp, Bhattacharyya:2008mz} for the result in
arbitrary dimensions). In particular, the various relaxation times
encountered at second order should descend to interesting transport
coefficients for the non-relativistic theory. It must however be
mentioned that unlike the relativistic case, where causality issues
\citep{Baier:2007ix} require us to consider non-linear hydrodyanamics,
one is not similarly forced to consider higher-order terms in the
non-relativistic setting. It would also be interesting to carry
out a systematic analysis of non-relativistic conformal hydrodynamics
by looking at the constraints on the allowed tensor structures coming
from the Schr\"odinger symmetry, paralleling the relativistic analyses
of \citep{Baier:2007ix, Loganayagam:2008is}.

The fluid dynamics we have discussed here has been restricted to
conformal fluids with Schr\"odinger symmetry. One main feature of such
fluids is that they are in general compressible; this follows from the
fact that the energy density is related to the pressure $2 \, \enr =
d\, P$ through the equation of state (which in turn follows from scale
invariance). To make contact with the usual studies of incompressible
Navier-Stokes equations we need to ensure that we can decouple the
fluctuations in the density. This can be achieved by looking at low
frequency modes which do not excite the propagating sound mode in a
hydrodynamic system. In fact, this limit was discussed recently in the
context of the fluid-gravity correspondence in
\citep{Fouxon:2008tb,Bhattacharyya:2008kq} where the authors showed
that starting from a parent relativistic conformal fluid dynamical
system one can recover incompressible Navier-Stokes equations in a
suitable scaling limit. Curiously the limit procedure reveals an
interesting structure in the fluid equations -- they are scale
invariant under a new scaling symmetry. This symmetry is different
from the Schr\"odinger symmetry enjoyed by the fluids under
consideration in this paper. We arrived at our hydrodynamic
description in $d$ spatial dimensions via a light-cone reduction of a
$d+2$ dimensional relativistic theory. In
\citep{Fouxon:2008tb,Bhattacharyya:2008kq} the authors derive
non-relativistic incompressible hydrodynamics in $d+1$ spatial
dimensions by a suitable limit of the relativistic theory. It would be
interesting to the take the incompressible limit of the
non-relativistic conformal fluid described here to compare with their
results.

\subsection*{Acknowledgments}
\label{acks}

It is a pleasure to thank Chris Herzog and Veronika Hubeny for
collaboration at initial stages of the project and for useful discussions.  
MR would like to thank ICTS, TIFR for hospitality
during the Monsoon Workshop in String theory and CERN for hospitality
during the Black Holes Theory Institute. MR and SFR are supported in
part by STFC.

\appendix
\section{Hamiltonian calculation in the bulk}
\label{appHam}

Since the field theory dual of the asymptotically Schr geometries is
non-relativistic, we might not expect the construction of a spacetime
stress tensor for asymptotically AdS geometries to have a natural
counterpart for these solutions. We might instead expect the map from
bulk to boundary to involve some kind of Hamiltonian framework, which
treats spatial and time directions separately. In this appendix, we
describe the construction of a Hamiltonian for the bulk geometry from
the action constructed in~\citep{Herzog:2008wg}. This is a
generalization of the calculation of \citep{Hawking:1995fd}, using the
known form of the action to determine the correct boundary terms for
the Hamiltonian framework.

The idea of the calculation in \citep{Hawking:1995fd} is to choose a
foliation of the spacetime by constant time slices, and rewrite the
covariant action in the form $\CS = \int dt\, [p \dot{q} - H]$, keeping
careful track of the boundary terms. This will then give us a form for
the Hamiltonian including boundary terms at infinity.  For simplicity
we start from the truncated version of the action used
in~\citep{Herzog:2008wg},\footnote{The analysis can be easily extended to the consistent truncation Lagrangian of \cite{Maldacena:2008wh} modulo the cost of some cumbersome expressions. We find that we have an eight parameter set of boundary terms which lead to the finite on-shell action reported in \cite{Herzog:2008wg}.}
\begin{eqnarray}
\CS &=& \frac{1}{16\pi\, G_5}\,\int d^5 x \sqrt{-g} \left(R 
- \frac{4}{3} (\partial_\mu \phi) (\partial^\mu \phi) 
- \frac{1}{4} e^{-8 \phi / 3} F_{\mu\nu} F^{\mu\nu} 
- 4 \,A_\mu A^\mu - V(\phi) \right)  \\ \nonumber
&&+ \frac{1}{16\pi G_5} \int d^4 \xi \,\sqrt{-h}\, \left(2\, K - 6 
+ A_\mu A^\mu + c \,A_\mu A^\mu \phi + b \,(A_\mu A^\mu)^2 + (2 \,c -4\,b +3) \phi^2\right)\ 
\label{finaction}
\end{eqnarray}
for some arbitrary constants $c, d$. To simplify the formulae below,
we will work with $c=b=0$, but it is straightforward to carry out the
analysis in the general case. We take the constant time slices to
be the surfaces $\Sigma_+$ of constant $x^+$. The coordinates on
$\Sigma_+$ will be denoted collectively as $\chi^\mu$ and the induced
metric is $\kappa_{\mu \nu}$.\footnote{We will not distinguish between the
  indices used for different hypersurfaces.} The lapse and shift $N$,
$N^\mu$ are defined by considering a vector $u^\mu$ defined on
$\Sigma_+$, such that $u^\mu \, \nabla_\mu x^+ = 1$, and decomposing this
vector into the normal $n^\mu $ to $\Sigma_+$ and the shift,
\begin{equation}
u^\mu = N \, n^\mu + N^\mu.
\end{equation}	
We also have a constant $r$ surface which is our cut-off on the
spacetime, with coordinates $\xi^\mu$ and induced metric $h_{\mu
  \nu}$. To top off the list of surfaces we have the constant time
surfaces on the cut-off surfaces which we will denote as
$\Sigma^\infty_+$ and we reserve $\zeta^\mu$ for the coordinates
and $\sigma_{\mu \nu}$ for the induced metric on $\Sigma^\infty_+$.

The analysis of the gravitational part of the action \req{finaction}
is very similar to the calculation in \citep{Hawking:1995fd}, and leads
to
\begin{equation}
H_{g} = \, \int_{\Sigma_+} d^4 \chi \, ( N \, \CH_g + N^\mu \, \CH_{\mu,g}) - \,
\int_{\Sigma^\infty_+} d^3 \zeta \sqrt{\sigma} \, 
\left[  \frac{N}{16 \pi G_5} (2\, ^{(3)}\! K
  -6) -2\,  N^\mu \, \frac{p_{\mu \nu}}{\sqrt{\kappa}} \, r^\nu \right],
\end{equation}	
where $p_{\mu \nu}$ is the
gravitational conjugate momentum, $p_{\mu\nu} = \sqrt{\kappa} (K_{\mu\nu} -
K g_{\mu\nu})$, and $r^\mu $ is the unit normal to
$\Sigma^\infty$ and $\kappa = {\rm det}(\kappa_{\mu \nu})$. The contributions to the bulk Hamiltonian and Hamiltonian density constraints are 
\begin{equation}
\CH_g = \frac{16 \pi\, G_5}{\sqrt{\kappa}} \, p_{\mu\nu} \,p^{\mu\nu} 
- \frac{16 \pi \,G_5}{3\,\sqrt{\kappa}} \,p^\mu_{\
  \mu} \,p^\nu_{\ \nu} - \frac{\sqrt{\kappa}}{16 \pi \,G_5} \, \CR, \quad \CH_{\mu,g}
= -2 \sqrt{\kappa}\, 
\CD_\nu \left( \frac{p_\mu^{\ \nu}}{\sqrt{\kappa}} \right). 
\end{equation}
Here $\CD_\mu$ is the covariant derivative associated with the
spatial metric $\kappa_{\mu\nu}$.
Only the boundary term will contribute to the on-shell Hamiltonian. We
note that this can be rewritten as
\begin{equation}
H_{g}^{os}  =  - \int\, d^3\zeta\, \left[ \sqrt{\sigma} \,
\frac{N}{16 \pi G_5} \left(2\, ^{(3)}\! K - 6 \, N\right) 
- 2\,N^\mu \, p_{\mu}^r \right].
\label{hamgos}
\end{equation}	

For the matter part of the action,
\begin{eqnarray}
\CS_m &=& \frac{1}{16 \pi\, G_5} \int d^5 x \,\sqrt{-g} \left(
  -\frac{4}{3}\, \partial_\mu \phi \,\partial^\mu \phi - \frac{1}{4}
  e^{-8\phi/3} F_{\mu\nu} F^{\mu\nu} - 4\, A_\mu A^\mu - V(\phi) \right)
\\ \nonumber
&&+ \frac{1}{16 \pi\, G_5} \int d^4 \xi \sqrt{-h} \,\left(
  A_\alpha A^\alpha + 3 \,\phi^2 \right),
\end{eqnarray}
performing the space-time split and introducing the momenta
\begin{equation}
p_\phi = \frac{1}{6 \pi \,G_5} \frac{\sqrt{\kappa}}{N} \,\left(\dot{\phi} -
N^\mu \partial_\mu \phi\right), 
\end{equation}
\begin{equation}
p_A^\lambda = \frac{e^{-8\phi/3}}{16 \pi\, G_5} \,\frac{\sqrt{\kappa}}{N} \,\left(\dot{A}_\mu
- \partial_\mu A_u - N^\nu F_{\nu\mu}\right) \kappa^{\lambda \mu},
\end{equation}
where dot denotes differentiation wrt $x^+$, we can rewrite this action as 
\begin{eqnarray}
\CS_m &=& \int dx^+ \int_{\Sigma_+} d^4 \chi \left[ p_\phi \dot\phi +
  p_A^\lambda \dot A_\lambda + \frac{2 \sqrt{\kappa}}{8 \pi \,G_5 \,N} (A_+ - N^\mu
  A_\mu)^2 - (\partial_\mu A_+) \,p_A^\mu - N \,\CH_m' - N^\mu \,\CH_{\mu,m}'
\right] \nonumber \\ &&+ \int dx^+ \int _{\Sigma^\infty_+} d^3 \zeta \sqrt{\sigma} \left[ -
  \frac{1}{16 \pi G_5 N} (A_+ - N^\alpha A_\alpha)^2
+
  \frac{N}{16 \pi G_5} \left( A_\alpha A^\alpha + 3 \,\phi^2 \right) \right].
\end{eqnarray}
We need to integrate the term $(\partial_\mu A_+)\, p_A^\mu$ by parts;
$A_+$ will then be a non-dynamical field, and we can eliminate it
using its equation of motion,
\begin{equation}
A_+ = N^\mu A_\mu - 2 \pi\, G_5 \,N\,  \CD_\mu \frac{p^\mu_A}{\sqrt{\kappa}},
\end{equation}
 This gives us the matter action in its
final form,
\begin{eqnarray}
\CS_m &=& \int dx^+ \int_{\Sigma_+} d^4 \chi \left[ p_\phi \,\dot\phi +
  p_A^\lambda \,\dot A_\lambda - N \,\CH_m - N^\mu \,\CH_{\mu,m}
\right] \\ \nonumber 
&&+ \int dx^+ \int_{\Sigma^\infty_+} d^3 \zeta \sqrt{\sigma} \left[ -
  \frac{1}{16 \pi \,G_5 \,N} (A_+ - N^\alpha A_\alpha)^2 - A_+
 \, \frac{p_A^\mu}{\sqrt{\kappa}} \,r_\mu \right. \\ \nonumber && \left. +
  \frac{N}{16 \pi\, G_5} \left(A_\alpha A^\alpha + 3 \,\phi^2 \right) \right], 
\end{eqnarray}
where
\begin{eqnarray}
\CH_m &=& \frac{8 \pi\, G_5 \sqrt{\kappa}}{8} \left(\CD_\mu
\frac{p_A^\mu}{\sqrt{\kappa}} \right)^2 + \frac{24 \pi \,G_5}{8\sqrt{\kappa}} p_\phi^2 +
\frac{8 \pi \,G_5}{\sqrt{\kappa}} \,e^{8\phi/3}\, 
p_A^\mu \,p_A^\nu \, 
\kappa_{\mu\nu} \\ \nonumber && + \frac{\sqrt{\kappa}}{16 \pi \,G_5} \left[ \frac{4}{3}\, \partial_\mu\phi \,\partial^\mu \phi + \frac{1}{4}\,  e^{-8\phi/3} \,F_{\mu\nu} F^{\mu\nu} + 4 \,A_\mu A^\mu + V(\phi) \right],
\end{eqnarray}
and
\begin{equation}
\CH_{\mu,m} = \partial_\mu \phi\, p_\phi + F_{\mu\nu}\, p_A^\nu -
A_\mu\, \sqrt{\kappa} \,\CD_\nu \frac{p_A^\nu}{\sqrt{\kappa}}. 
\end{equation}
Thus, the matter contribution to the on-shell Hamiltonian is
\begin{equation}
H_{m}^{os}  =  - \int_{\Sigma^\infty_+}\, d^3\zeta\, 
  \sqrt{\sigma} \,
  \left[ -
  \frac{1}{16 \pi \,G_5 \,N} (A_+ - N^\alpha A_\alpha)^2 - A_+\,
  \frac{p_A^\mu}{\sqrt{\kappa}}\, r_\mu  +
  \frac{N}{16 \pi\, G_5} \left( A_\alpha A^\alpha + 3 \,\phi^2 \right) \right].
\end{equation}
On-shell
\begin{equation}
A_+ - N^\alpha A_\alpha \propto A^+ = 0,
\end{equation}
so we can drop the first term, which will not contribute to the value
or first variation of the Hamiltonian, and write 
\begin{equation}\label{Hmat} 
  H_{m}^{os} = - \int_{\Sigma^\infty_+}\, d^3\zeta\,
  \, \left[ - A_+\, p_A^r + \frac{N\sqrt{\sigma}}{16 \pi\, G_5} \left(
      A_\alpha A^\alpha + 3 \,\phi^2 \right) \right],
\end{equation}
which is closer in form to the gravitational part.

For our spacetime \req{5dbh}, we have (writing for brevity $\ell(r) = \sqrt{f(r)}\, k(r)^{1/6}$)
\begin{eqnarray}
u^\mu &=& \frac{\partial}{\partial x^+}, \nonumber \\
n^\mu &=& -\frac{\gamma}{r^3\, \ell(r)} \,  
\frac{\partial}{\partial x^+} - \frac{r\,(1+f(r))}{2\, \gamma\, \ell(r)}\,  \frac{\partial}{\partial x^-},
\end{eqnarray}	
which leads to 
\begin{eqnarray}
N &=& -\frac{1}{\gamma}\, r^3\, \ell(r), \nonumber \\
N^\mu &=& -r^4 \frac{1+f(r)}{2\,\gamma^2}\, \frac{\partial}{\partial x^-},
\end{eqnarray}	
and $r^\mu$ being the unit-normal to $\Sigma^\infty_+$ is just given as 
\begin{equation}
r^\mu = \frac{r\, f(r)}{\ell(r)} \, \frac{\partial}{\partial r}.
\end{equation}	
Direct computation gives 
\begin{eqnarray}
\sqrt{\tilde h} &=& \gamma\, r, \nonumber \\
^{(3)}\!K &=& \frac{f(r)}{\ell(r)} \ , \nonumber \\
\frac{p_{-r}}{\sqrt{k}} &=& - \frac{N}{8 \pi \,G_5}\;
\frac{\gamma^2}{r^5 \,f(r) \,k(r)}. 
\end{eqnarray}	
The gravitational part of the Hamiltonian then involves
\begin{eqnarray}
H_{g}^{os}  &=&  - \int_{\Sigma^\infty_+}\, d^3\zeta\, \sqrt{\sigma} \,
\frac{N}{16 \pi \,G_5} \left( 2\,\frac{f(r)^{1/2}}{k(r)^{1/6}} - 6 + 2
\,  k(r)^{-7/6} \frac{(1+f(r))}{f(r)^{1/2}} \right) \\ \nonumber
&\approx& - \int_{\Sigma^\infty_+}\, d^3\zeta\, \sqrt{\sigma} \,
\frac{N}{16 \,\pi G_5} \left( -5 \,\frac{\gamma^2}{r^2} + \frac{21}{4}\,
  \frac{\gamma^4}{r^4} - \frac{r_+^4}{r^4} \right).
\end{eqnarray}
We see that the $r^4$ divergence cancels. The $r^2$ divergence will
cancel against a contribution from the matter part, which we turn to
next.

On the 3-boundary,
\begin{equation}
A_\alpha A^\alpha = \frac{\gamma^2}{r^2\, k(r)^{4/3}}, 
\end{equation}
so
\begin{equation}
A_\alpha A^\alpha +3 \,\phi^2  \approx \frac{\gamma^2}{r^2} - \frac{7}{12}
\,    \frac{\gamma^4}{r^4}. 
\end{equation}
We also evaluate
\begin{equation}
-A_+ \frac{p_A^r}{\sqrt{\kappa}} r_r = \frac{e^{-8 \phi/3}}{8 \pi\, G_5\, N} \,\frac{r
  f(r)^{1/2}}{k(r)^{1/6}} \,A_+ (\partial_r A_+ -
N^- \partial_r A_-) \approx \frac{N}{16 \pi \,G_5} \left( \frac{4
    \gamma^2}{r^2} - \frac{14}{3}\, \frac{\gamma^4}{r^4} \right)
\end{equation}
to obtain
\begin{equation}
H_{m}^{os} \approx - \int_{\Sigma^\infty_+}\, d^3\zeta\, \sqrt{\sigma} \,
\frac{N}{16 \pi \,G_5} \left( 5\, \frac{\gamma^2}{r^2} - \frac{21}{4}\,
  \frac{\gamma^4}{r^4} \right). 
\end{equation}

As a result, the total Hamiltonian is
\begin{equation}
H^{os} = \frac{1}{16 \pi \,G_5} \int_{\Sigma^\infty_+}\, d^3\zeta\, \sqrt{\sigma} \, N \, \frac{r_+^4}{r^4} = -\frac{1}{16 \pi\, G_5} \int_{\Sigma^\infty_+}\, d^3\zeta\, r_+^4, 
\end{equation}
in agreement with our expectations: this matches the total energy
obtained by Euclidean methods in \citep{Herzog:2008wg}. 

It is easy to also extract the particle number density from this
calculation; we simply consider shifting the vector $t^\mu \to t^\mu +
\alpha^\mu$, where $\alpha^\mu \partial_\mu t = 0$, which shifts
$N^\mu \to N^\mu + \alpha^\mu$. This will change 
\begin{equation}
p_A^\lambda \to p_A^\lambda - \frac{e^{-8\phi/3} \, \sqrt{\kappa}}{16 \pi \,G_5 N}
\alpha^\nu \, F_{\nu \mu} \,k^{\lambda \mu},
\end{equation}
and hence redefines the Hamiltonian by
\begin{equation}
H^{os} \to H^{os} - \int_{\Sigma^\infty_+}\, d^3\zeta\, \sqrt{\sigma}\, \left[ -2 \,\alpha^\mu \,\frac{p_{\mu\nu}}{\sqrt{\kappa}} \,r^\nu 
+ \frac{e^{-8\phi/3}}{16 \pi\, G_5 \,N}
\alpha^\nu \,F_{\nu \mu} \,r^\mu - \frac{1}{16 \pi\, G_5 N}
(\alpha^\nu A_\nu)^2 \right]. 
\end{equation}
We should interpret this new Hamiltonian associated with $t^\mu +
\alpha^\mu$ as a combination of the energy and the momentum in the
direction specified by $\alpha^\mu$. We see immediately that if
$\alpha^\nu$ points in one of the spacelike directions, the change in
the Hamiltonian vanishes, which agrees with our expectation that the
spatial momentum densities vanish in the state we are considering. If
we take $\alpha^\nu = \alpha \,\delta^\nu_-$, then the gravity term is
\begin{equation}
\frac{p_{-r}}{\sqrt{\kappa}}\, r^r = 
- \frac{N}{8 \pi\, G_5} \;\frac{\gamma^2}{r^5 \,f(r)\, k(r)}
\frac{r \,f(r)}{\ell(r)} \approx - \frac{N}{8 \pi \,G_5}\,
  \frac{\gamma^2}{r^4}. 
\end{equation}
The matter terms give
\begin{equation}
\frac{e^{-8\phi/3}}{16 \pi \,G_5 \,N}
\alpha^\nu \,F_{\nu \mu} \,r^\mu \approx - \frac{\alpha N}{8 \pi \,G_5}
\frac{\gamma^4}{r^6}, \quad  \frac{1}{16 \pi \,G_5 \,N}\,
(\alpha^\nu A_\nu)^2 \approx \frac{\alpha^2 \,N}{16 \pi \,G_5}\;
\frac{\gamma^6}{r^{10}},
\end{equation}
so only the gravitational term contributes to the change in the
Hamiltonian, which is
\begin{equation}
H^{os} \to H^{os} - \int_{\Sigma^\infty_+}\, d^3\zeta\, \sqrt{\sigma}\, N \frac{1}{4 \pi \,G_5} \;\frac{\gamma^2}{r^4} = H^{os} 
- \int_{\Sigma^\infty_+}\,d^3\zeta\, \frac{1}{4 \pi\, G_5} \gamma^2,
\end{equation}
consistent with the value $\langle P_- \rangle 
= \gamma^2 V \Delta x^-/4 \pi G_5$
obtained in the Euclidean approach.

Thus, this approach gives us a definition of the Hamiltonian whose
on-shell value appears correct. We should ask if the Hamiltonian so
defined is the generator of the asymptotic time-translation symmetry
on this class of spacetimes. To be the generator of the symmetry, our
Hamiltonian should satisfy~\citep{Regge:1974zd}
\begin{equation} \label{hamcond}
\delta H = \int d^4 \chi \left[ \delta p_{\mu\nu} \,E^{\mu\nu} + \delta
p_\psi \,E^\psi + \delta k^{\mu\nu} \,A_{\mu\nu} + \delta \psi\, A_\psi\right],
\end{equation}
with no surface terms, where $\psi$ denotes the collection of matter
fields. Now, our Hamiltonian constructed from the covariant action
should satisfy this property by construction, since \req{hamcond} is
just the requirement for Hamilton's principle $\delta \int dt\, (p
\dot{q} - H) =0$ to have solutions, and the boundary terms in our
action $\CS =\int dt \,(p \dot{q} - H)$ were chosen precisely so as to
ensure that $\delta \CS =0$ on-shell. However,
in~\citep{Herzog:2008wg}, the action was only shown to satisfy $\delta
\CS =0$ for a restricted set of boundary conditions on the variations,
relating the leading order variations of different fields. It would
therefore be useful to check directly that the Hamiltonian we have
proposed satisfies \req{hamcond}. We have encountered difficulties in
performing this check; we leave their resolution to future work.


\providecommand{\href}[2]{#2}\begingroup\raggedright\endgroup

\end{document}